\documentclass[amsmath,amssymb,letterpaper]{revtex4}

\usepackage[verbose]{wrapfig}
\usepackage{subfigure}
\usepackage{amsmath,amssymb}
\usepackage{xcolor}
\usepackage[final]{graphicx}

\newcommand{\bvec}[1] {{\mathbf {#1}}}
\newcommand{\Einc} { { E_\text{inc} } }

\DeclareMathOperator{\sech}{sech}

\makeatletter
\begin{document}

\title{Simulations of the nonlinear Helmholtz equation: arrest of beam collapse, nonparaxial solitons and counter-propagating beams}

\author{{G. Baruch}$^{1*}$, {G. Fibich}$^{1}$ and {Semyon Tsynkov}$^{2}$ }

~

\address{
	$^1$School of Mathematical Sciences, Tel Aviv University, Tel Aviv 69978,
	Israel\\
	$^2$Dep. of Mathematics, North Carolina State University, Box 8205, 
	Raleigh, NC 27695, USA\\
	$^*$Corresponding author, guy.baruch@math.tau.ac.il
}

%
\begin{abstract}
We solve the $\left.(2+1)D\right.$ nonlinear Helmholtz equation (NLH) for input
beams that collapse in the simpler NLS model. Thereby, we provide the first ever
numerical evidence  that nonparaxiality and backscattering can arrest the
collapse.
We also solve the $\left.(1+1)D\right.$ NLH and show that solitons with radius
of only half the wavelength can propagate over forty diffraction lengths with no
distortions.
In both cases we calculate the backscattered field, which has not been done
previously.
Finally, we compute the dynamics of counter-propagating solitons using the NLH
model, which is more comprehensive than the previously used coupled NLS model.
\end{abstract}

\maketitle

The nonlinear Schr\"odinger equation (NLS) is the canonical model in nonlinear optics for propagation of intense laser beams in isotropic Kerr media.
In the case of propagation through a bulk medium, Kelley~\cite{Kelley:65} used
the $2D$ NLS to predict the possibility of a catastrophic collapse of beams
whose input power is above the {\em critical power for collapse}. 
In the case of propagation through planar waveguides, the 1D NLS was used to predict the existence of spatial solitons \cite{Askaryan:62}.
Both beam collapse in bulk medium and spatial solitons in planar waveguides were
observed in experiments~\cite{Shen:75, Barthelemy-Maneuf-Froehly:85}.
More recently, configurations of two counter-propagating beams were modeled by two coupled NLS equations~\cite{Cohen-counterprop:2002}.

In nonlinear optics, the NLS is derived from the nonlinear Maxwell equations via
a series of approximations. 
First, if the electric field is monochromatic and third harmonic generation is
neglected, Maxwell's equations reduce to the vector nonlinear Helmholtz equation
(NLH).
If the field is also linearly polarized, the vector NLH reduces to the 
scalar NLH~\cite{Fibich-Ilan-Vectorial-PhysicaD}. 
Finally, the NLS is derived from the  scalar NLH using the paraxial approximation, which is valid when the beam radius is sufficiently large compared with the wavelength.
As, however, the $2D$ NLS predicts that the beam  radius shrinks to zero at collapse, the paraxial approximation breaks down at this point.
In the case of spatial $1D$ solitons, the paraxial approximation sets a lower limit on the soliton radius.

The singular behavior of the $2D$ NLS solutions for collapsing beams is non-physical.
Therefore, an important question is whether the singularity formation is already arrested by taking one step back in the aforementioned series of approximations and employing the scalar NLH model, or only in a more comprehensive model. 
Both the mathematical analysis and simulations of the scalar NLH have proved to be considerably more difficult than for the NLS, since for the NLH  one solves a nonlinear boundary-value problem, whereas the NLS requires solving an initial value problem.
An additional computational obstacle is that unlike the NLS, which governs the slowly varying envelope, the NLH has to be approximated with sub-wavelength resolution.
For these reasons, the question of collapse in the scalar NLH model was not fully answered for over 40 years.

Previously, numerical simulations and asymptotic 
analysis~\cite{Vlasov:87, Feit-Fleck, Fibich-PRL:96} suggested that
nonparaxiality arrests the collapse in bulk medium.
These studies, however, applied various simplifying approximations to the scalar
NLH.  In particular, they considered only forward traveling waves and completely neglected the backscattered field. 
Even though backscattering is generally believed to be ``small'', it may still significant affect the overall propagation, because collapse dynamics in the $2D$ cubic NLS is extremely sensitive to small perturbations~\cite{Fibich-Papanicolau:99}.

\maketitle

To study the arrest of collapse in the scalar NLH with no simplifying assumptions (and in particular, with the backscattering included), Fibich and Tsynkov developed a fixed-point iterative numerical method for solving the NLH as genuine boundary value problem \cite{FT:01, FT:04}, which is based on freezing the nonlinearity at each iteration.
This method converged for input  powers below the critical power for collapse $P_{cr}$, but diverged for input powers higher than $P_{cr}$.
It was unclear, however, whether the divergence above $P_{cr}$ was due to limitations of the numerical method, or because collapse is not arrested in the scalar NLH model.
Subsequently, the method of~\cite{FT:01, FT:04} was used to show
numerically the arrest of collapse in the linearly-damped scalar 
NLH~\cite{damped:03}.
In that work, however, the magnitude of damping was much larger than in actual physical settings, and could not be reduced to zero.
More recently, Sever proved existence of solutions (and hence arrest of 
collapse) in the scalar NLH  with self-adjoint boundary 
conditions~\cite{sever:06}.
The proof in~\cite{sever:06}, however, relies heavily on self-adjointness, 
whereas propagating fields satisfy radiation boundary conditions (BCs), which
are non self-adjoint.
Therefore, until now, there has been no conclusive evidence that the collapse is arrested in the scalar NLH model.

In~\cite{BFT:07}, we studied numerically the $\left.(0+1)D\right.$ NLH, which models the propagation of plane waves in a Kerr medium.
In this case, the solution always exists, but becomes non-unique (bistable) above a certain input power threshold~\cite{chen-mills-PRB:87}.
Numerically, we observed that the fixed-point frozen nonlinearity method of~\cite{FT:01, FT:04}  converges for low input powers, but diverges for higher powers which are still below the threshold for non-uniqueness.
This indicates that the divergence of the fixed-point frozen nonlinearity method is due to the numerical methodology itself, rather than to non-uniqueness or non-existence of the solutions.
Therefore, an alternative iterative solver, based on Newton's method, was constructed and shown to have much better convergence properties.
In this Letter, we extend the Newton-based method of~\cite{BFT:07} to the multi-dimensional case.
The resulting technique enables us to solve the $\left.(2+1)D\right.$ NLH  for input powers above $P_{cr}$.
Hence, we obtain the first ever computational evidence that the  collapse of the beam is indeed arrested in the scalar NLH model.
We also calculate the field backscattered from the domain.
Moreover, we solve the $\left.(1+1)D\right.$ NLH for a ``nonparaxial'' soliton 
with radius equal to half a wavelength, and observe that it propagates virtually 
unchanged over $40$ diffraction lengths. This indicates that such beams are 
still in the paraxial regime.
Finally, we solve the $\left.(1+1)D\right.$ NLH for two counter-propagating beams and compare the results to those obtained using the coupled NLS model.

\begin{figure}
	\vspace{10pt}
	\begin{center}
	\fbox{ 
		\includegraphics[clip,scale=0.15]{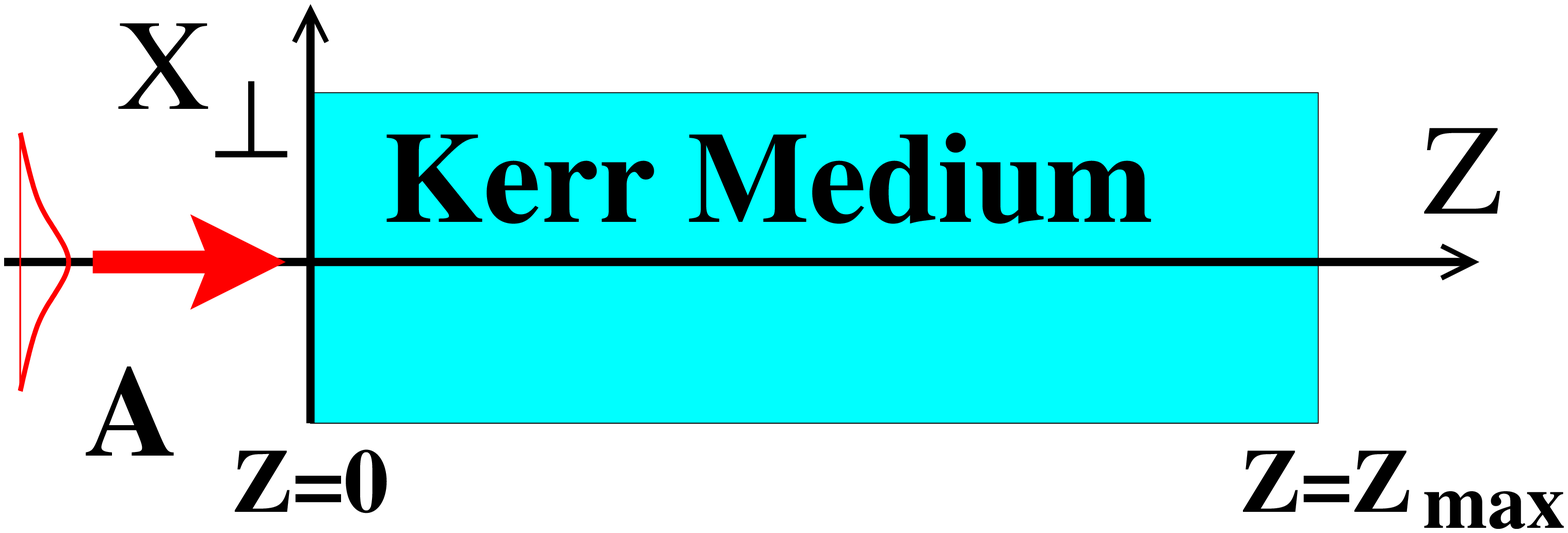}
	}
	\fbox{ 
		\includegraphics[clip,scale=0.145]{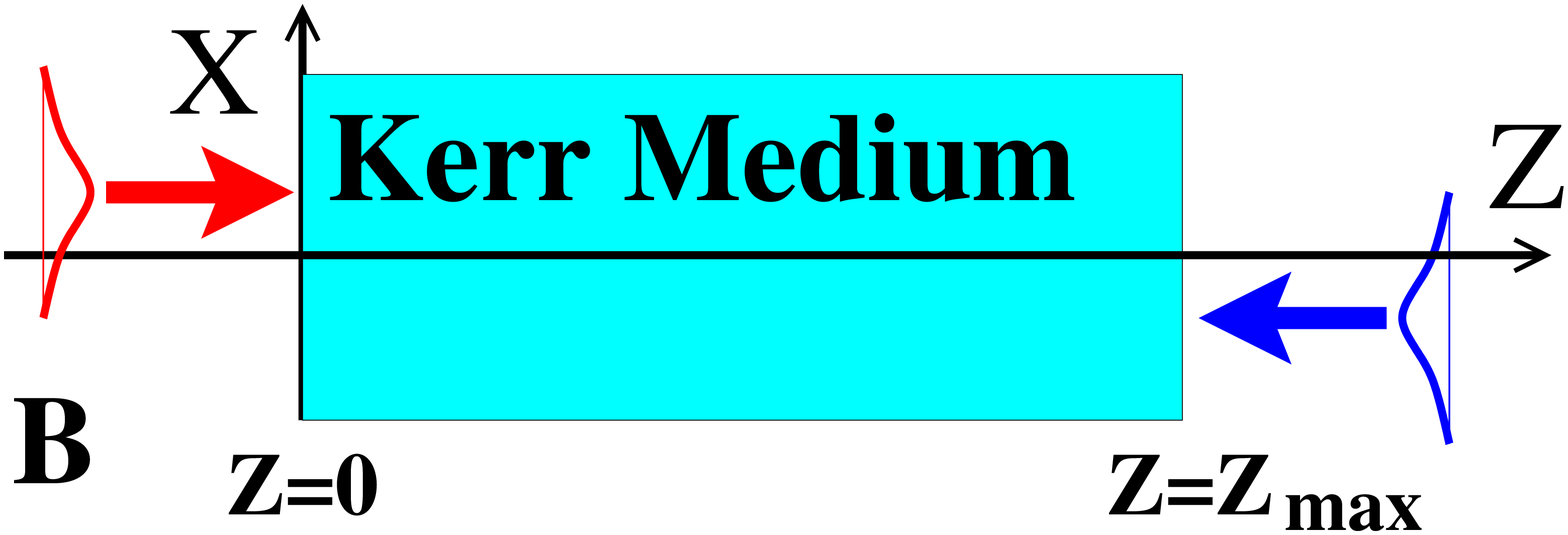}
	}
	\fbox{ 
		\includegraphics[clip,scale=0.175]{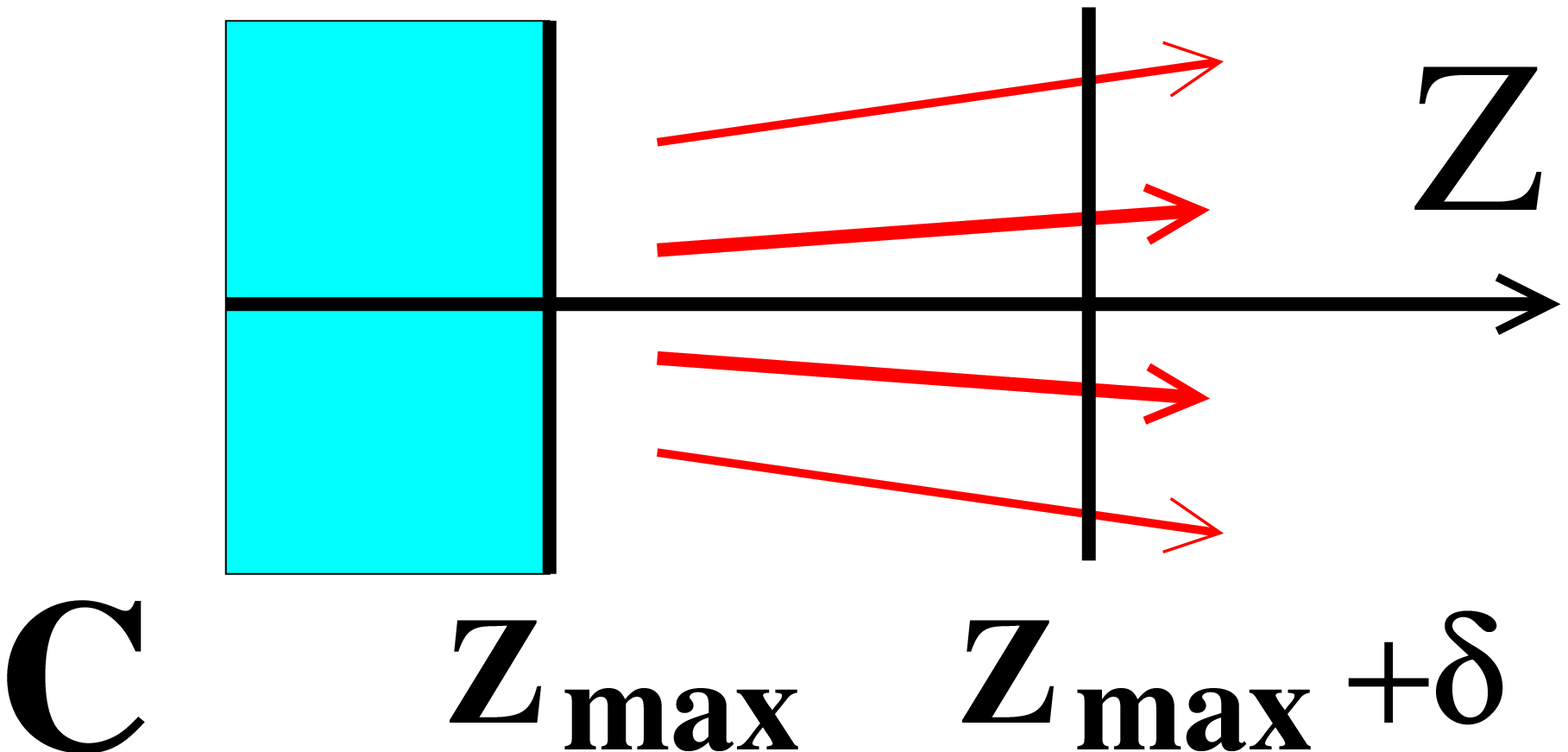}
	}
	\fbox{ 
		\includegraphics[clip,scale=0.18]{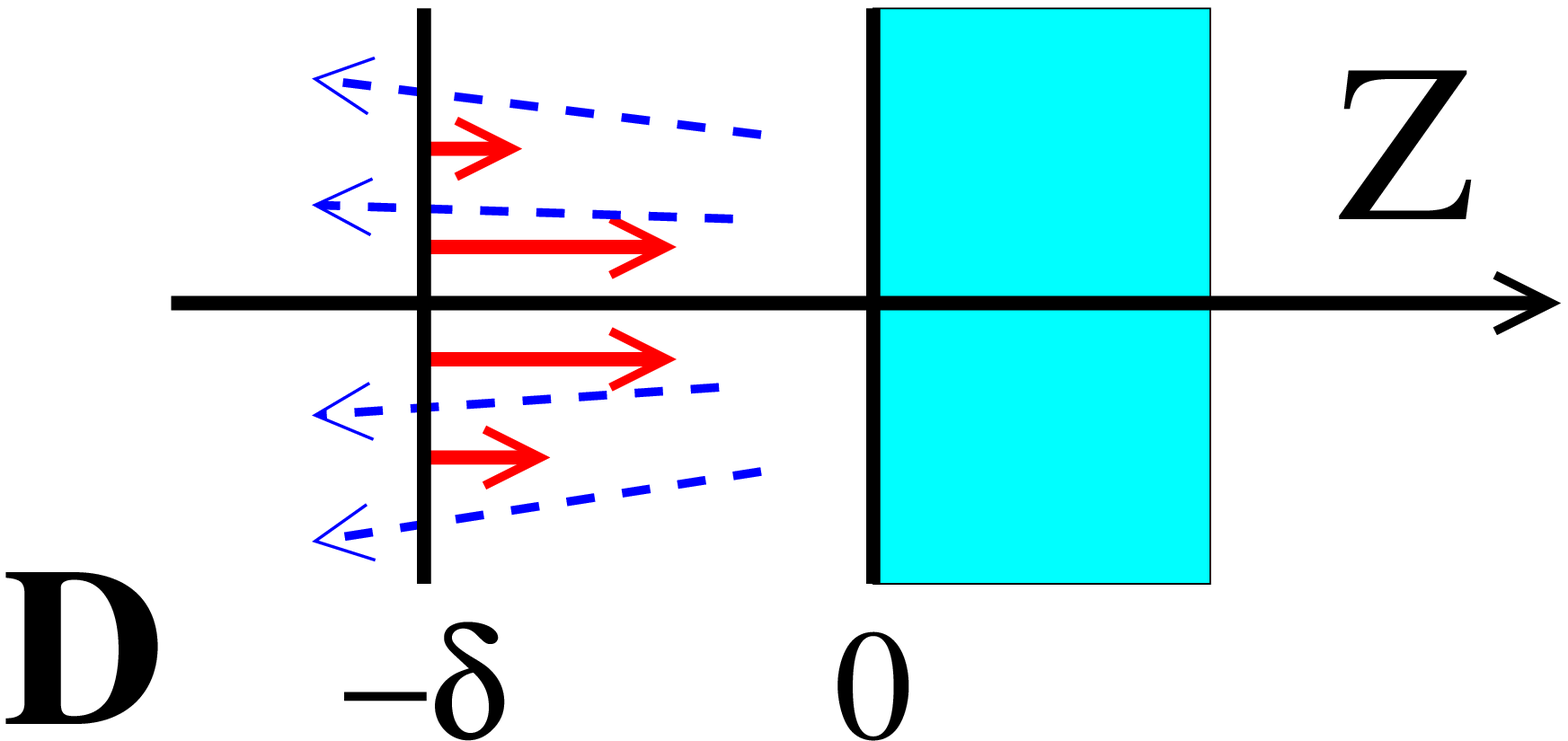}
	}
	\caption{\label{fig:setup} The physical setup: 
		A:~single beam,
		B:~counter-propagating beams.
		C:~A schematic of the upstream BC at $z=Z_{\max}+\delta$, which freely
		admits all forward propagating waves (red).
		D:~A schematic of the downstream BC at $z=-\delta$, which freely admits
		all the backward propagating waves (blue) to pass, and also specifies
		the (forward moving) incoming beam.
	}
\end{center}\end{figure}
The propagation of linearly polarized, continuous wave beams in isotropic Kerr media is governed by the scalar nonlinear Helmholtz equation:
\begin{equation}
	E_{zz}(z,\bvec{x}_\perp) + \Delta_\perp E + k_0^2
 \left(1+(2n_2/n_0)|E|^2\right) E = 0,
	\label{eq:NLH} 
\end{equation}
where 
	$E$ is the electric field,
	$k_0$ is the linear wavenumber,
	$n_0$ is the linear index of refraction 
	and $n_2$ is the Kerr coefficient.
In the bulk medium $\left.(2+1)D\right.$ case $\left.\bvec{x}_\perp=(x,y)\right.$ and $\left.\Delta_\perp=\partial_x^2+\partial_y^2\right.$;
in the planar waveguide $\left.(1+1)D\right.$ case $\left.\bvec{x}_\perp=x\right.$ and $\left.\Delta_\perp=\partial_x^2\right.$.
We consider an incoming beam traveling in the positive $z$ direction (henceforth
``forward'' or ``right'') impinging on a finite-length Kerr material slab at the
$z=0$ interface and exiting the Kerr medium at the $z=Z_{\max}$ interface, see
Fig.~\ref{fig:setup}(A).
A portion of the field may be reflected by the interfaces at $z=0$ or
$z=Z_{\max}$, or backscattered inside the Kerr medium, because of the variations
of the index of refraction induced by the forward-propagating beam.
To derive the NLS, the standard approach is to represent the field as
$E=Ae^{ik_0z}$, where the envelope $A$ is assumed  slowly varying.
Using the standard rescaling 
	$\bvec{\tilde{x}}_\perp=\bvec{x}_\perp/r_0$,
	$\tilde{z}=z/2L_{DF}$
	and $
		\tilde{A}(\tilde{z},\bvec{\tilde{x}}_\perp)
		=\sqrt{2n_2/n_0}r_0k_0\cdot A(z,\bvec{x}_\perp),
	$ where $r_0$ is the input beam radius and $L_{DF}=k_0r_0^2$ is the diffraction length, the NLH can be written in the dimensionless form 
\begin{equation}	
	f^2 \tilde{A}_{\tilde{z}\tilde{z}}(\tilde{z},\bvec{\tilde{x}}_\perp) 
	+ i \tilde{A}_{\tilde{z}} + \Delta_\perp \tilde{A} 
	+ |\tilde{A}|^2\tilde{A}=0, \label{eq:NLH-dimensionless}
\end{equation}
where $
	f^2=(r_0k_0)^{-2}=\left.(\frac{\lambda_0}{2\pi r_0})^2\right.
$ is the {\em nonparaxiality parameter}.
Typically  $\lambda_0 \ll r_0$ so that $f^2\ll 1$ and $f^2 \tilde{A}_{\tilde{z}\tilde{z}}\ll \tilde{A}_{\tilde{z}}$.
Therefore, the paraxial approximation, which consists of neglecting $f^2 \tilde{A}_{\tilde{z}\tilde{z}}$, leads to the NLS
\begin{equation}	
	i \tilde{A}_{\tilde{z}}(\tilde{z},\bvec{\tilde{x}}_\perp) 
	+ \Delta_\perp \tilde{A} + |\tilde{A}|^2\tilde{A}=0. \label{eq:NLS}
\end{equation}

In our simulations, the $\left.(2+1)D\right.$ NLH with cylindrical symmetry,
i.e., $E=E(z,r)$ where $
	r=\left|\bvec{x}_{\perp}\right|=\sqrt{x^2+y^2}
$, is approximated with a fourth order finite-difference scheme.
The solution is computed for $-\delta \leq z \leq Z_{\max}+\delta $
in order to implement the BCs in the linear regions.
At the material interfaces $z=0$ and $z=Z_{\max}$ where the index of refraction is discontinuous, Maxwell equations for a normal-incident 
field imply that $E$ and $E_z$ are continuous across the interfaces.
At $z=Z_{\max}+\delta$ we imposed the radiation BC that the field does not have
any left-going component for $z>Z_{\max}$, see Fig.~\ref{fig:setup}C.
Similarly, at $z=-\delta$ we implement the two-way radiation BC that for $z<0$
the field does not have right-going components except for the prescribed
incoming beam which impinges on the interface $z=0$ with a transverse
profile~$\Einc(r)$, see Fig.~\ref{fig:setup}D.
Because $z=-\delta$ and $z=Z_{\max}+\delta$ are outside the Kerr slab, the field 
propagation there is linear, which simplifies the implementation of the
radiation BCs, see~\cite{FT:01, FT:04} for more details.
The discretized system of nonlinear algebraic equations is solved using Newton's method~\cite{BFT:07}.

In order to focus on the effects of the Kerr nonlinearity, the values of $n_0$
in the Kerr medium ($0\leq z\leq Z_{\max}$) and in the surrounding linear medium
($z<0$ and at $z>Z_{\max}$) are chosen to be equal, so that to eliminate the
reflections due to discontinuity of $n_0$ at the interfaces.
However, discontinuities in the nonlinear coefficient are not eliminated, 
and are a source of reflections at $z=0, Z_{\max}$.
Our numerical method can be applied to the case of different $n_0$ 
 with no change~\cite{BFT:07}.
Note that, since we solve the NLH in non-dimensional form, {\em simulations in
this Letter are valid for any physical value of $k_0, n_0,n_2$} that corresponds
to the same dimensionless quantities $f^2$ and $P/P_{cr}$.

\begin{figure}
	$
		\begin{array}{c}
	\text{ \fbox{A
		\includegraphics[clip,scale=0.7]{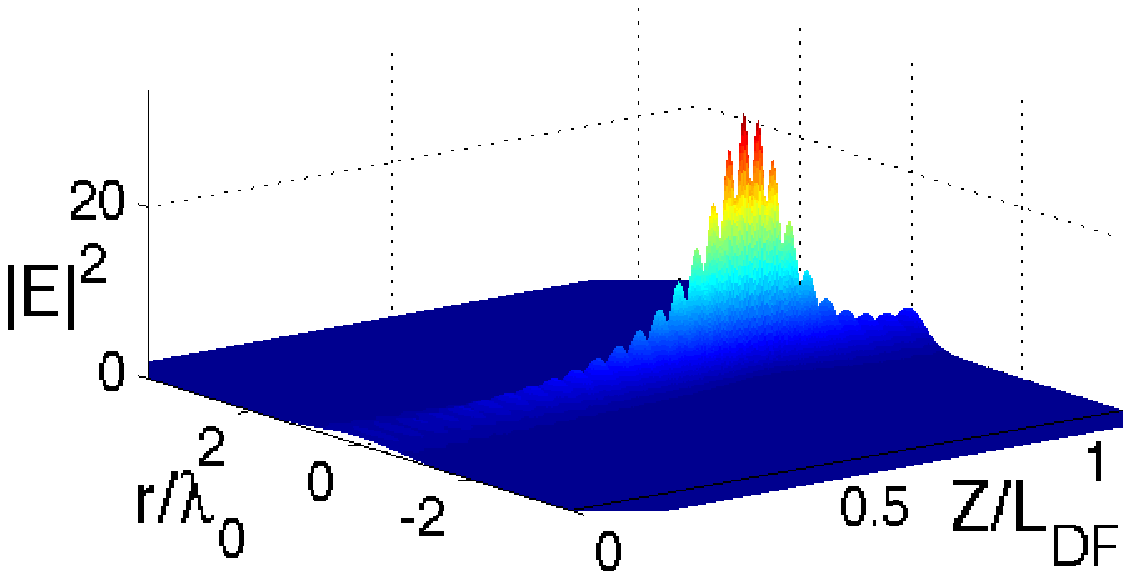}
		\includegraphics[clip,scale=0.6]{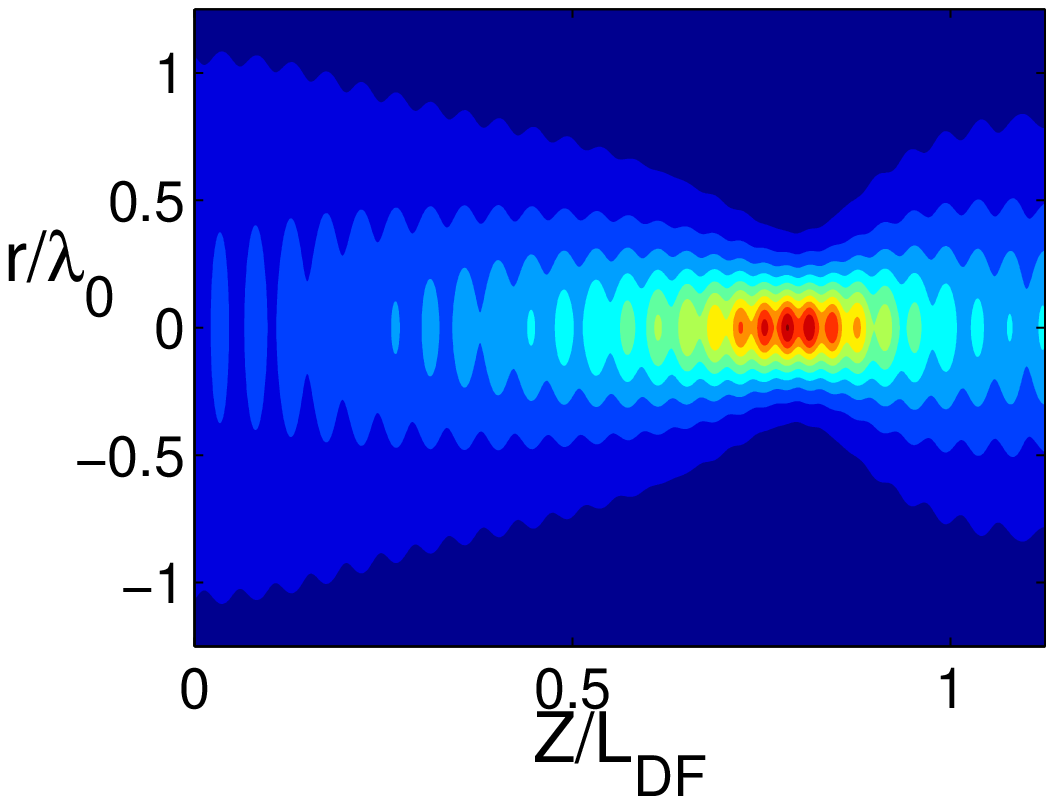}}
	}\\
	\text{\fbox{B
		\includegraphics[clip,scale=0.7]{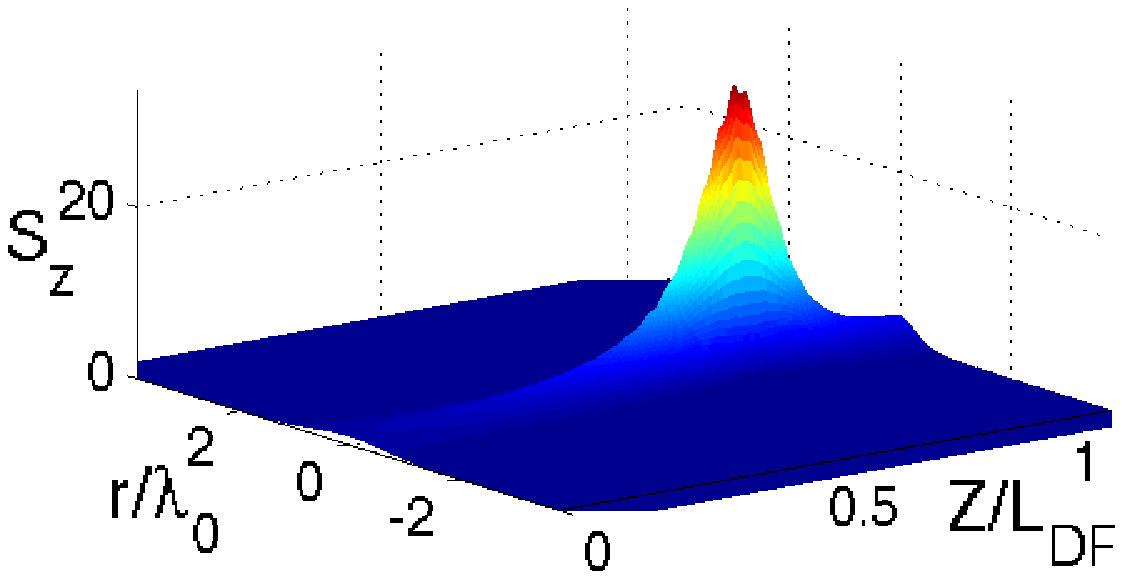}
		\includegraphics[clip,scale=0.6]{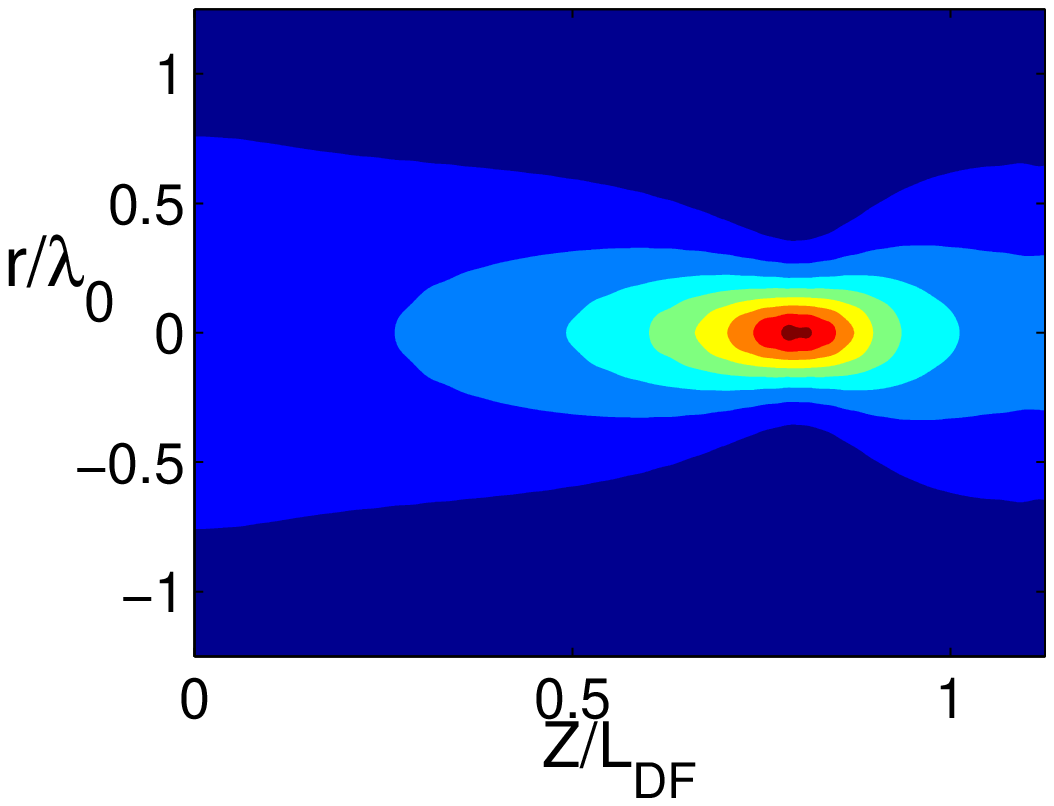}}
	}
		\end{array} 
	$
	\caption{(color online)\label{fig:collapse} Arrest of collapse in 
		the $\left.(2+1)D\right.$ NLH.
		A:~$|E|^2$. 
		B:~$S_z$.
	}
\end{figure}

\begin{wrapfigure}{r}{0.4\textwidth}
	\begin{center}
		\includegraphics[clip,width=0.35\textwidth]{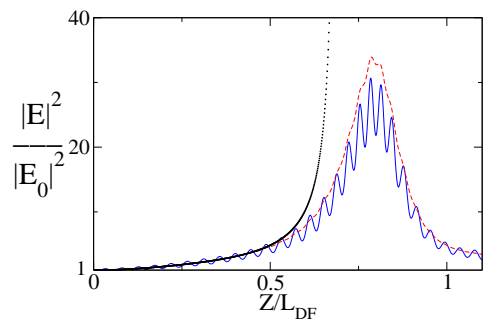}
		\caption{\label{fig:collapse-oa}
			Arrest of collapse in the $\left.(2+1)D\right.$ NLH: 
			comparison of normalized on-axis~$|E|^2$ (blue solid),
			$S_z$~(red dashed), and NLS solution (black dotted).
		}
	\end{center}
\end{wrapfigure}

The $\left.(2+1)D\right.$ NLH (\ref{eq:NLH}) was solved for an incoming collimated Gaussian beam $
	\Einc=\left.(\sqrt{2n_2/n_0}r_0k_0)^{-1}\right.
	e^{-(r/r_0)^2}
$ of radius $r_0=1.27 \lambda_0$, corresponding to nonparaxiality parameter of $f^2=(k_0r_0)^{-2}=1/64$, and input power of $P=1.29 P_{cr}$.
The NLH solution initially self-focuses, until $z\approx 0.8L_{DF}$ where the
collapse is arrested, after which the solution defocuses, see
Fig.~\ref{fig:collapse}(A).
The corresponding NLS solution collapses at $z_c=0.68L_{DF}$, see
Fig.~\ref{fig:collapse-oa}.
This comparison of the NLH and NLS provides a direct numerical evidence that {\em collapse is arrested in the scalar NLH model}.

The fast oscillations of $|E|^2$ in the $z$ direction in
Fig.~\ref{fig:collapse}(A) are not a numerical artifact, but rather account 
for the actual physics.
Indeed, let us first note that a part of the forward-propagating wave is reflected backwards by the material interfaces at $z=0$ and  $z=Z_{\max}$.
In addition, since the forward propagating beam induces changes in the
refraction index, part of the beam may be backscattered inside the Kerr medium.
The presence of both forward and backward traveling fields, i.e,
\begin{equation}
	\label{eq:coupled_AB}
	E\approx Ae^{ik_0z}+Be^{-ik_0z},
\end{equation}
implies that $
	|E|^2\approx |A|^2+|B|^2+2\text{Re}\left(AB^*e^{i2k_0z}\right).
$
Hence, $|E|^2$ should undergo oscillations with wavenumber $\sim2k_0$. 
Note that the analytical solutions of the $\left.(0+1)D\right.$ NLH also exhibit these $2k_0$ intensity oscillations~\cite{chen-mills-PRB:87}.
The prediction that the intensity undergoes $2k_0$ oscillations implies that the index of refraction also oscillates.
In other words, the backward traveling field induces a $2k_0$ Bragg grating.
This prediction may be tested by pump-probe experiments.

In order to find a smoother representation of the solution, 
recall that for the NLS~(\ref{eq:NLS}) the conserved beam power is $
	P_{NLS}=\int|\tilde{A}|^2d\bvec{\tilde{x}}_\perp.
$
For the NLH~(\ref{eq:NLH}), however, the conserved beam power is $
	P_{NLH}=\int S_z d\bvec{x}_\perp,
$ where $
	\bvec{S}=k_0\text{Im}(E^*\nabla E)
$ is the energy flux, or Poynting vector, and $
	S_z=k_0\text{Im}\left.(E^* \frac{\partial E}{\partial z})\right.
$ is its $z$-component.
Specifically, for the field~(\ref{eq:coupled_AB}) the value of $S_z$ reduces to the flux difference $S_z\approx k_0^2\left(|A|^2-|B|^2\right)$.
It is therefore much smoother than $|E|^2$, and provides a ``more natural''
depiction of the NLH solution, as confirmed by comparing $S_z$ of
Fig.~\ref{fig:collapse}(B) with $|E|^2$ of Fig.~\ref{fig:collapse}(A).
The energy flux $S_z$ shows the arrest of collapse and the focusing-defocusing
dynamics more clearly, see also Fig.~\ref{fig:collapse-oa}.

\begin{figure}[t]
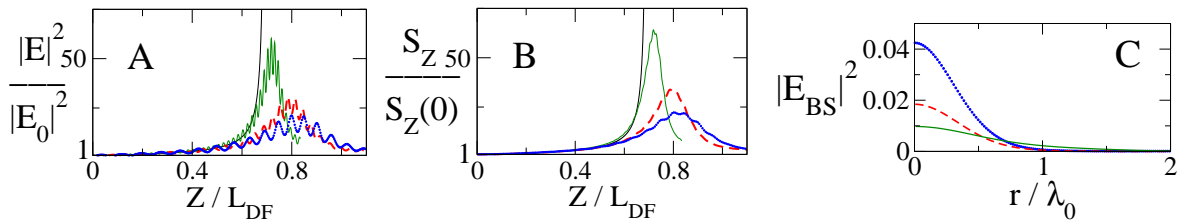


	~\\

	\begin{center}
	\includegraphics[clip,width=0.55\textwidth]{3ks_combined.eps}
	~
	\includegraphics[clip,width=0.3\textwidth]{3ks_E2_BS.eps}
	\caption{\label{fig:3ks_collapse} (color online)
		NLH solutions with 
		$r_0/\lambda_0=\frac{3}{\pi}$ (blue, dots), 
		$\frac{4}{\pi}$ (red, dash)
		and $\frac{6}{\pi}$ (green, solid).
		Solid black line is the NLS solution.
		A:~Normalized on-axis intensity $|E/E(z=0)|^2$.
		B:~Normalized on-axis Poynting flux $S_z/S_z(z=0)$.
		C:~Transverse profile of the backward field at $z=0-$.
	}
	\end{center}
\end{figure}

In order to analyze the effect of the nonparaxiality parameter $f^2$, in
Fig.~\ref{fig:3ks_collapse} we fix the wavelength and vary the input beam radius
$r_0$ (while keeping the power unchanged) so that $f^{-2} = 36,\ 64$, and $144$.
All the NLH solutions initially follow the collapsing NLS solution, but later 
the collapse is arrested and the solution defocuses.
As expected, for a wider input beam (lower nonparaxiality), the deviations from the NLS solution and the arrest of collapse occur later, and the maximum self-focusing is higher.
Again we see that $|E|^2$ has $2k_0$ oscillations (whose magnitude increases as the input beam becomes more nonparaxial), while the energy flux $S_z$ is smooth.

Our numerical algorithm for solving the NLH also enables the computation of backscattering from the Kerr medium.
In Fig.~\ref{fig:3ks_collapse}(C), we present the backward propagating field 
profiles for the previous three solutions, just before the material interface 
at $z=0$.
To the best of our knowledge, {\em this is the first ever calculation of the
backscattered field of collapsing beams,}
which is due to backscattering from inside the Kerr medium and reflections from
the nonlinear interface.
As the input beam radius $r_0$ decreases, the power of the backscattered field increases from $0.46\%$ to $0.63\%,$ to $2.1\%$ of the incoming beam power.
This, as well as a comparison of magnitudes of oscillations in
Fig.~\ref{fig:3ks_collapse}(A), shows that the backscattered field increases as the input beam becomes more nonparaxial.

\begin{figure}
	\begin{center}
	\subfigure{
		\fbox{
			\includegraphics[clip,scale=0.6]{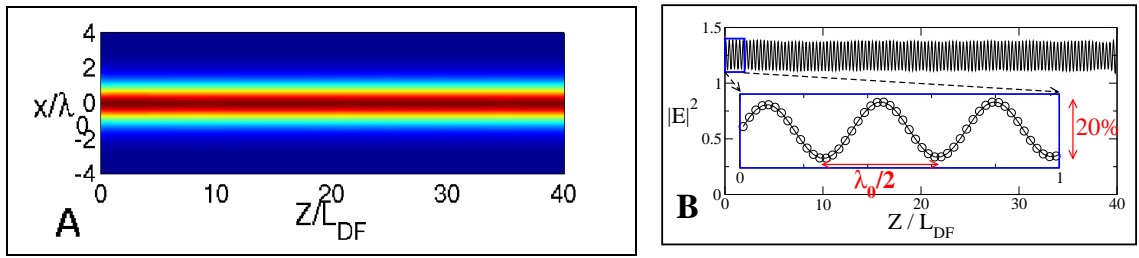} 
		}
	}
	\subfigure{
		\includegraphics[clip,scale=0.25]{single_sol_axis_E2_160x30_2000x150_e0.2_k2.eps}
	}
	\end{center}
	\caption{ \label{fig:single_sol} (1+1)D NLH soliton 
		with $r_0=\frac{\lambda_0}{2}$ propagating over $40L_{DF}$.
		A: $S_z$.
		B: On-axis $|E|^2$.
	}
\end{figure}
In $\left.(1+1)D\right.$ configurations, the NLS possesses stable soliton
solutions.
It is generally believed that the paraxial approximation breaks down when the
beam width becomes comparable to $\lambda_0$ and that, therefore, no solitons of
such narrow width exist.
To see that this is not the case, the $\left.(1+1)D\right.$ NLH (\ref{eq:NLH})
is solved 
for the incoming NLS-soliton profile $
	\Einc=\left.(\sqrt{2n_2/n_0}r_0k_0)^{-1}\right. \sech(x/r_0)
$ with width $r_0=\lambda_0/2$, 
impinging on a Kerr slab of finite length $Z_{\max}=40 L_{DF}$.
As in the $(2+1)D$ case, we impose continuity of $E$ and $E_z$ at the material
interfaces $z=0$ and $z=Z_{\max}$, and apply the radiation BCs in the linear
regions at~$z=-\delta$ and $z=Z_{\max}+\delta$.
The solution inside the Kerr-slab resembles a ``nonparaxial~soliton'' which
propagates virtually unchanged, see Fig.~\ref{fig:single_sol}(A).
We note that even for such a narrow beam, the nonparaxiality parameter is still
moderate, as $f^2=1/\pi^2\approx0.1$, which may explain why there there still
exist soliton-like solutions.
Similarly to the $\left.(2+1)D\right.$ case, because a part of the forward
propagating beam is backscattered, $|E|^2$ exhibits the fast $2k_0$ oscillations
(Fig.~\ref{fig:single_sol}(B)), while $S_z$ is smooth.
In this case, the backscattered field leads to $10\%$ oscillations in $|E|^2$.

Posada, McDonald and New \cite{Posada-McDonald-New:1, Posada-McDonald-New:2}
studied solutions of the $\left.(1+1)D\right.$ Helmholtz
equation over a semi-infinite Kerr medium, of the form $
	A(z,x) =
		\left.(\sqrt{2n_2/n_0}r_0k_0)^{-1}\right.
		\sech(x/r_0) \cdot e^{i c\cdot z}.
$
In later works they found similar stationary states for different nonlinearities
\cite{Posada-McDonald-New:3, Posada-McDonald-New:4}.
These solutions do not have any backward propagating components.
In contrast, for the finite-length Kerr medium simulation of
Fig~\ref{fig:single_sol}, some backward moving waves must exist, because of
reflections from the material discontinuity at~$z=Z_{\max}$, and the full NLH as
a boundary-value problem must be solved.

\begin{figure} \begin{center}
	
	~\\

	\fbox{A
		\includegraphics[clip,scale=0.6]{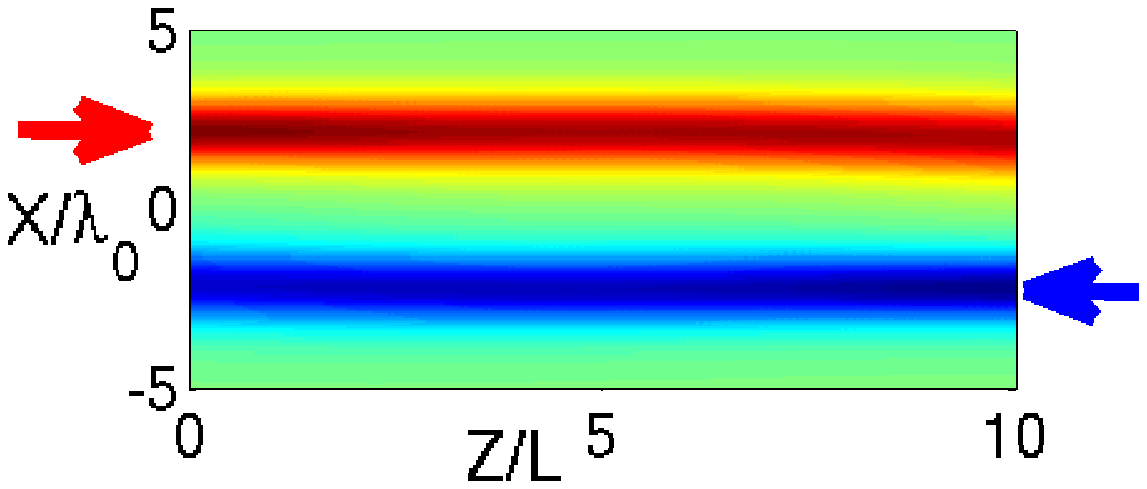}
	}
	\fbox{B
		\includegraphics[clip,scale=0.6]{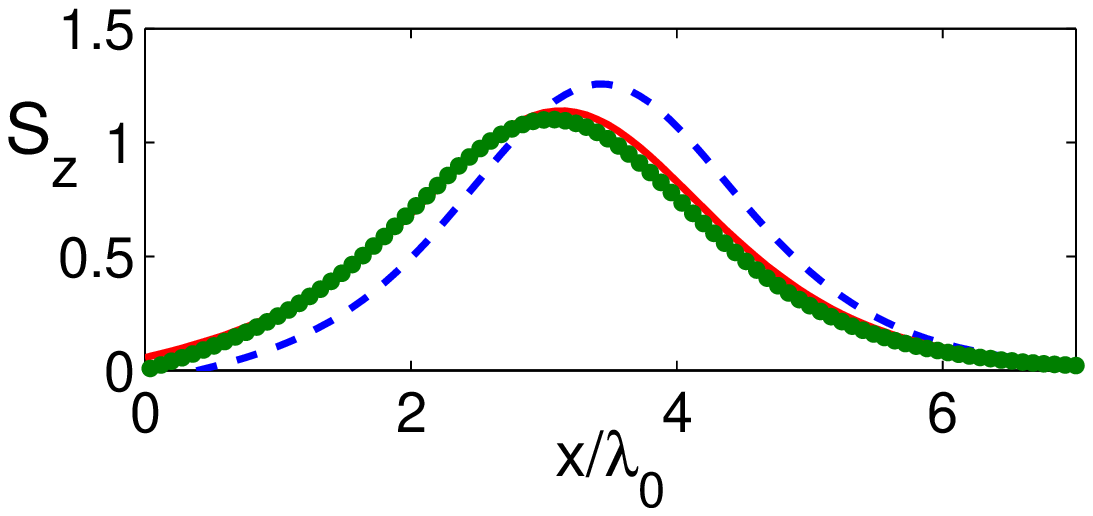}
	}
	\caption{\label{fig:cntr-prop}
		(color online)
		Energy flux ($S_z$) of the $\left.(1+1)D\right.$ NLH with counter-propagating beams.
		A: Positive (forward) flux is red, negative (backward) flux is blue.
		B: The right-going beam at its incoming (blue dashed) and outgoing interface (red solid).
		Green dotted line is the coupled-NLS solution at the outgoing interface.
	}
\end{center}\end{figure}

Another $\left.(1+1)D\right.$ configuration of recent interest is that of
counter-propagating beams, when a right traveling  soliton impinges at the left
interface and a left traveling beam impinges at the right interface
(Fig.~\ref{fig:setup}(B)).
This configuration was analyzed numerically by Cohen 
et al.\cite{Cohen-counterprop:2002} using a coupled NLS system, which is
derived from the NLH by employing the paraxial approximation and further
assuming that asynchronous terms of the Kerr nonlinearity can be neglected.
In doing so, the BCs should simultaneously account for the coupled incoming and outgoing fields at each interface. 
As noted in \cite{Cohen-counterprop:2002}, these BCs can only be approximately accommodated in the coupled NLS model.
In contrast, they can be fully implemented in the NLH model, without any
approximation. 
Fig.~\ref{fig:cntr-prop} presents our solution of the NLH for
counter-propagating beams of radius $r_0=\lambda_0$ that enter a Kerr material
slab at the opposite interfaces with a transverse displacement of
$d=4.4\lambda_0$, and propagate over $10L_{DF}$.
It shows that the beams are slightly attracted toward each other and also become
wider as they propagate.
The results are in close agreement with the coupled NLS model, see
Fig.~\ref{fig:cntr-prop}(B). 
Therefore, the more comprehensive NLH model confirms the validity of the coupled NLS model for counter-propagating beams even for the ``extreme'' parameters of $r_0=\lambda_0$ and $d=4.4\lambda_0$.

In this work, we solve the scalar NLH~(\ref{eq:NLH}), which is the simplest
model for the propagation of light in Kerr media that incorporates nonparaxial
effects, backscattering and reflections from material interfaces.
Moreover, unlike the NLS, it accurately models the propagation of oblique beams
and reflections from interfaces at arbitrary angles.
This model neglects vectorial effects, i.e., linear and nonlinear coupling
between the three components of the electric field.
We note that the vectorial effects scale as $f^2$, and hence are of the same
order of magnitude as nonparaxiality.
In bulk media, they have been shown to have the same effect as nonparaxiality,
which is to arrest the 
collapse~\cite{Chi-Gou:95, Fibich-Ilan-Vectorial-PhysicaD}.
In contrast, in planar waveguides, the solitons are stable.
Hence, when $r_0=\lambda_0/2$, $f^2$ is small, and so we expect that both 
nonparaxial and vectorial effects are likely to have a secondary effect on 
the propagation dynamics. 
Therefore, the sub-wavelength solitons predicted in the Letter are likely to remain stable also in the more comprehensive vector NLH model.  

The research of G.B. and G.F. was partially supported by Grant No.~2006-262 from
the US--Israel Binational Science Foundation (BSF). The work of S.T. was
supported by the US NSF, Grant No.~DMS-0509695, and US Air Force, Grant
No.~FA9550-07-1-0170.


\end{document}